\documentclass[twocolumn,runningheads]{svjour2}
\smartqed  
\usepackage{graphicx}
\journalname{Astrophysics and Space Science}
\begin{document}

\title{\bf Observations of AGNs using PACT 
}


\author{D. Bose         \and
        V. R. Chitnis    \and
 P. R. Vishwanath \and
 P. Majumdar \and \\
M. A. Rahman \and
B. B. Singh \and
        A. C. Gupta       \and 
        B. S. Acharya 
}


\institute{Tata Institue of Fundamental Research \\
           Homi Bhabha Road, Colaba, Mumbai 400 005, India\\
              \email{D. Bose $<$debanjan@tifr.res.in$>$}             \\
              \email{V. R. Chitnis $<$vchitnis@tifr.res.in$>$} \\
             \emph{Present addresses: } \\ 
             P. R. Vishwanath \at 
         Indian Institute of Astrophysics, Bangalore 560 034, India\\
              A. C. Gupta \at
               Yunnan Astronomical Observatory, Kunming, Yunnan, 650011 China P. R. \\   
              P. Majumdar \at
               Max-Planck-Institute for Physics, Foehringer Ring 6, 80805 Munich, Germany 
}

\date{Received: date / Accepted: date}

\maketitle

\begin{abstract}

    We report our observations on 4 AGNs viz, Mkn421, Mkn501,
    1ES1426+428 and ON231 belonging to a sub-class called
    blazars. The observations were carried out using the Pachmarhi
    Array of Cherenkov Telescopes and span about 6 years
    period from 2000 to 2005. We discuss our methods of analysis
    adopted to extract the gamma ray signal from cosmic ray
    background. We present our results on the emission of TeV
    gamma rays from these objects. Also, we report on the status
    of the new  high altitude experiment, a 7 telescope array at Hanle
    in the Himalayas at an altitude of about 4200 m above mean sea
    level for detecting celestial gamma rays.

\keywords{Mkn 421 \and Blazar \and $\gamma$-ray astronomy}
\end{abstract}

\section{Introduction}
\label{intro}
AGNs have dominated extragalactic $\gamma$-ray astronomy by virtue of their
great luminosities. The general understanding is that a supermassive 
black hole of mass $10^{6}-10^{9}$ M$_\odot$ at the center of AGN 
accretes mass from the surrounding medium, forming an accretion disk and
two jets emanating perpendicular to the plane of the accretion disk \cite{Ref1}. 
These jets channel a
plasma flowing out with relativistic speed and any radiation produced inside
them is greatly modified by Doppler effect. AGNs with jets directed towards us
are called blazars. Blazars are characterized by two distinct parts in their 
Spectral 
Energy Distributions (SEDs). First part in SED rises smoothly from radio 
wavelengths upto a broad peak spanning the range from optical to X-ray 
wavelengths and is due to relativistic electrons radiating via synchrotron 
process. Second part is probably due to inverse compton scattering of 
synchrotron photons by the same electrons and is characterized by a peak in 
SEDs in hard X-ray to $\gamma$-ray band \cite{Ref2}. One of the characteristic 
features of these blazars is their time variability on scales ranging from 
minutes to years. Mkn 421, Mkn 501, 1ES1426+428, ON231 are the four blazars 
observed using Pachmarhi Array of Cherenkov Telescopes (PACT) during 2000 to 
2005. In the following  sections we present our observations, analysis procedure 
and the
results obtained. Also, we have attempted a comparative study of two states 
of Mkn 421 namely, flaring and quiscent states, using radio, optical, X-ray 
and $\gamma$-ray data. 
Our future plans for the observations of $\gamma$-ray 
sources using a high altitude array are also presented. 

\section{Pachmarhi Array of Cherenkov Telescopes} 
\label{sec:1}

\begin{figure}
\centering
  \includegraphics[width=0.4\textwidth]{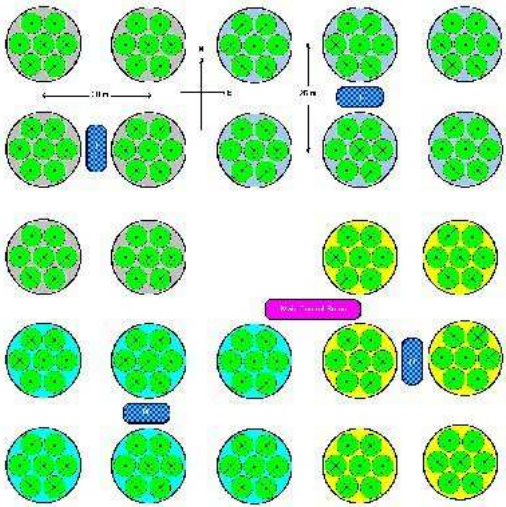}
\caption{A Layout of PACT. The big circles represent the telescopes. Seven smaller circles inside a big circle represent 7 mirrors in a telescope. Rectangular boxes represent the data acquisition centers.}
\label{fig:1}
\end{figure}

\begin{figure}
\centering
  \includegraphics[width=0.4\textwidth]{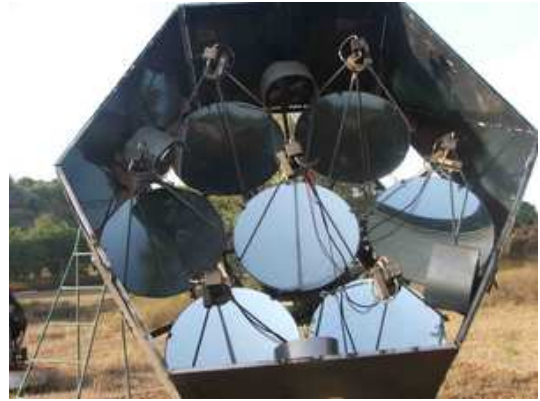}
\caption{Picture of one telescope with seven parabolic mirrors and PMTs}
\label{fig:2}
\end{figure}

    Pachmarhi Array of Cherenkov Telescopes (PACT) is located in Central
    India (latitude 22$^\circ$ 28$^\prime$ N, longitude 78$^\circ$ 25$^\prime$
    E, altitude 1075 $m$). We use {\it wavefront sampling technique} to
    detect {\it TeV} $\gamma$-rays from astronomical sources. There are 24
    telescopes spread over an area of $80m \times 100m$. Figure \ref{fig:1}
    shows the schematic layout of PACT. Each telescope has 7
    para-axially mounted parabolic mirrors of diameter 0.9m with a PMT
    (EMI 9807B) at the focus of each mirror as shown in figure \ref{fig:2}. 
    Entire array is sub-divided into 4 sectors with 6 telescopes in each. 
    Each sector has its own
    data acquisition system (DAQ) where data on real time, relative arrival 
    time of PMT
    pulses (using TDCs) and photon density (using ADCs) of six peripheral
    mirrors in a telescope are recorded.
    Apart from this, there is also a Master DAQ at the center of the array
    for recording information of an event relevant to entire array.
    PMT pulses of 7 mirrors in a telescope are linearly added to form a
    telescope pulse for trigger generation. Data recording is initiated when a
    coincidence of 4 out of 6 telescope pulses generates an event trigger for a
    sector. The typical trigger rate was about 2-3 Hz per sector. The orientations 
    of telescopes are controlled remotely and monitored
    throughout the observations \cite{Ref3}. Details of the setup can be 
    found elsewhere \cite{Ref4,Ref5}. To estimate energy threshold,
    collection area etc, we have carried out Monte Carlo simulations of extensive
    air showers using CORSIKA package developed by
    KASKADE group \cite{Ref6}. Energy threshold of PACT is
    estimated to be 750 GeV for vertically incident showers initiated by 
    $\gamma$-rays\footnote{Assumed energy spectrum F(E)=kE$^{-2.4}$dE} and the 
    corresponding collection area is 1.38$\times$10$^{5}$m$^{2}$. 
    The flux level for detecting $\gamma$-rays from a source at 5$\sigma$ 
    sensitivity  in 50 hours of observation using PACT is estimated to be 
     2.93$\times$10$^{-11}$  {\it ph cm$^{-2}$ s$^{-1}$} assuming no cosmic 
    ray rejection.
    In terms of Crab nebula flux this corresponds to about 0.9 Crab
    units above energy threshold of 750 GeV. 
    For inclined showers, simulations were carried out using IACT 
    option in CORSIKA. The energy threshold and collection area increases 
    with incident angle and these parameters are summarized in table \ref{tab:1}. 

\begin{table}[t]
\caption{Energy thresholds and collection areas for PACT at different incident\ angles}
\centering
\label{tab:1}
\begin{tabular}{lll}
\hline\noalign{\smallskip}
Inclination  &   Energy  &  Collection \\
  Angle      &  Threshold&    Area     \\
             &   (TeV)   &     (m$^2$)    \\ \hline
\tableheadseprule\noalign{\smallskip}
Vertical (0$^\circ$)     &    0.75 &  1.4$\times$10$^5$ \\
15$^\circ$   &    0.90 &  1.5$\times$10$^5$ \\
30$^\circ$   &    1.20 &  1.8$\times$10$^5$ \\
45$^\circ$   &    2.20 &  2.7$\times$10$^5$ \\
\noalign{\smallskip}\hline
\end{tabular}
\end{table}

\section{Observations and Data Analysis}
\label{sec:2}
  Observations were carried out using PACT on clear moon less nights.
Observations on source (source runs) were usually taken by pointing
all telescopes to the source direction. The typical run span was about 1-3
hours. Background runs were taken either immediately before or after 
(sometimes both before and after) the source run by
aligning all telescopes to a dark region
(a region with the same declination as that of the source but with different 
RA). Background region is chosen in such a way that it covers same zenith angle
range as that of the source. The observation log is given in table \ref{tab:2}.
For each source an equal amount of data were collected on corresponding 
background runs.

\begin{table}[t]
\caption{Observation log for blazars}
\centering
\label{tab:2}
\begin{tabular}{lllll}
\hline
\emph{Year} &
\multicolumn{4}{c}{\emph{Duration of Observations (minutes)}} \\\cline{2-5} 
\hline\noalign{\smallskip}
           & Mkn 421 & Mkn 501 & ON 231 & 1ES1426+428 \\
            & z=0.030 & z=0.034 & z=0.102 & z=0.129\\ \hline
\tableheadseprule\noalign{\smallskip}
    2000     &  3510.  &   710.  & --    &   --   \\
   2001     &  1960.  &   --    & --    &   --   \\
   2002     &  1860.  &   510.  & --    &  1520. \\
   2003     &  1770.  &   840.  &  510. &   570. \\
   2004     &  2270.  &   780.  &  550. &   870. \\
   2005     &   930.  &   --    & --    &   960. \\
\noalign{\smallskip}\hline
\end{tabular}
\end{table}

\begin{figure}
\centering
  \includegraphics[width=0.4\textwidth]{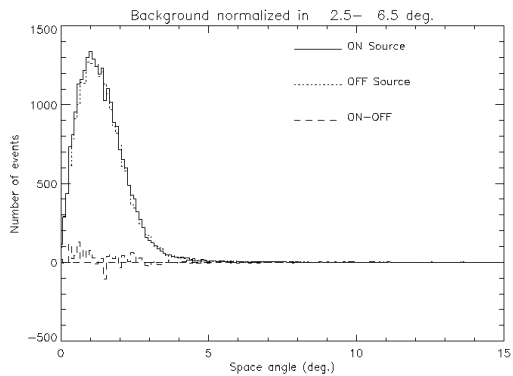}
\caption{Typical space-angle distribution of a source run (solid), background 
 run (dotted) and the difference between the two (dashed).}
\label{fig:3}
\end{figure}

    PACT data were analysed in the following way.
 Celestial $\gamma$-rays are not affected by the
 interstellar magnetic field, therefore they retain their directionality.
 Whereas cosmic rays, being charged particles, are scattered by the
 interstellar magnetic field, as a result they are isotropic. Thus a source
 emitting $\gamma$-rays will be reflected as an excess of events from the
 source direction compared to off-source direction. 
 A number of preliminary checks were carried out on the data before doing 
 actual analysis. A cut is imposed on the number of telescopes with valid 
 TDC data to be $\geq$ 8. Care is taken to see that the distribution of 
 telescopes with
 valid TDC data are similar in both source and background data sets.
 The arrival direction of each shower is determined by reconstructing
 shower front using the relative arrival times of Cherenkov photons at
 various telescopes (or PMTs). Cherenkov photon front is then fitted with
 a plane, normal to this plane gives the direction of the shower axis. Then, 
 for each shower or event, the space angle is estimated as an angle between the
 direction of shower axis and the source direction. Thus space angles are
 obtained for all events in source as
 well as background runs. Space angle distributions of all source runs
 are compared with the corresponding distributions
 of background runs over the same ${\it zenith}$ angle coverage. Figure 
\ref{fig:3} shows the space angle distributions of events from source and 
 background runs  taken in a night. Space angle distribution of background 
 events is normalised to the source distribution by comparing the shape of 
 the distributions in 2.5$^\circ$ to 6.5$^\circ$ window since we do not 
 expect any $\gamma$-ray event in this region~\cite{Ref5}. This normalisation 
 is necessary since there are
 variations in the sky conditions at different times of the same night.
 Differences between the number of source and background
 events is calculated for each bin as ($S_{i}$-c$B_{i}$) where c is a constant.
 We define, $$\chi^2=\sum_{i=2.5}^{6.5}  (S_{i}-cB_{i}){^2}$$
 and normalisation constant $c$ is chosen such that $\chi^2$ is
 minimum. The difference between the source and normalised background events
 in 0$^\circ$ to 2.5$^\circ$ region is then used as $\gamma$-ray signal. 
  Thus time averaged $\gamma$-ray signal is obtained for each night observations. 

\subsection{Analysis of Multiwavelength Data of Mkn 421}
\label{sec:3}

   We have attempted a comparative study of high and low states of Mkn 421.
  We chose a flare in March/April 2001 as representative of high state 
  as this flare is one of the strongest flares and February/March 2003 as a 
  representative of low state.  
   We have analysed X-ray archival data from Proportional Counter Array (PCA) 
 on board RXTE obtained during these two periods. The first 
 data set corresponds to the period 19th March to 1st April 2001 when Mkn421 was in 
 flaring state. 
 The second set was collected during 26th February to 5th March 2003, Mkn421 
 was in quiescent state at this time.
 PCA data were extracted from archival data sets of RXTE satellite\footnote{http://heasarc.gsfc.nasa.gov/W3Browse} 
 from observation ids 60145 and 80172 respectively. 

\begin{figure}
\centering
  \includegraphics[width=0.5\textwidth]{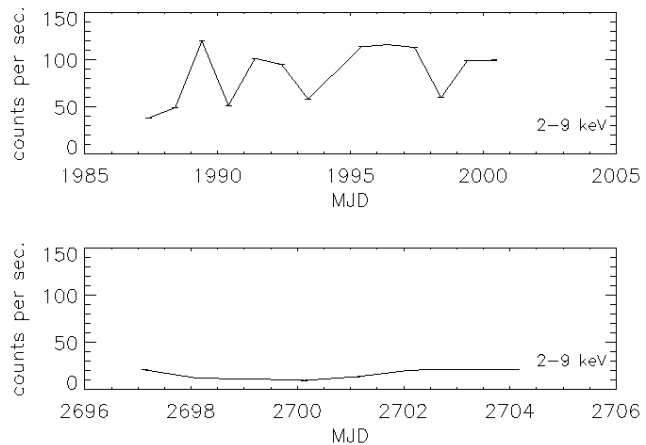}
\caption{2-9 keV light curve obtained from PCA data, upper panel for 2001 and
lower panel for 2003.}
\label{fig:4}
\end{figure}

 The PCA consists of five identical xenon
 filled proportional counter units (PCUs) covering an energy range of 2-60 keV.
 During these observations only PCU0 and PCU2 were used. We have analyzed
 Standard 2 PCA data which has a time resolution of 16s with energy information
 in 128 channels. Even though observations were carried out with PCU 0 and
 PCU 2, we have used only PCU 2 data because PCU 0 had lost its front
 veto layer at the beginning of year 2000. So the data from PCU0 are 
 more prone to
 contamination by events caused by low-energy electrons entering the detector.
 Data reduction is done with FTOOLS (version 5.3.1)\footnote{see
 http://heasarc.gsfc.nasa.gov/docs/software/lheasoft} distributed as part of
 HEASOFT (version 5.3). For each of the observations, data were filtered using
 standard procedure given in the RXTE Cook Book
 \footnote{http://heasarc.gsfc.nasa.gov/docs/xte/recipes/cook\_book.html}.
 For extraction of background, model appropriate for bright sources\footnote{
 (pca\_bkgd\_cmbrightvle\_eMv20031123.mdl)} was used for the 2001 data since
 the source was in high state during that period. For 2003 data, when the 
 source was in low state, the model appropriate for faint sources\footnote{
 (pca\_bkgd\_cmfaintl7\_eMv20030330.mdl)} was used. 
  The resulting X-ray light curves for the two states of Mkn421 are shown in 
 figure~\ref{fig:4}.

 Contemporaneous optical/NIR and radio data are also available for Mkn421 
 corresponding to the two data sets mentioned above. Optical data in V-band 
 for year 2001 was taken by WEBT
(Whole Earth Blazar Telescope) using KVA-telescope on La Palma \cite{Ref7}.
 For 2003, Near Infra Red data in J-band was taken at Gurushikhar observatory,
 Mount Abu \cite{Ref8}. Radio data was taken by Mets$\ddot{a}$hovi radio
 telescope at 22 GHz and 37 GHz \cite{Ref9}.

\section{Results}
\label{sec:4}

\begin{figure}
\centering
  \includegraphics[width=0.5\textwidth]{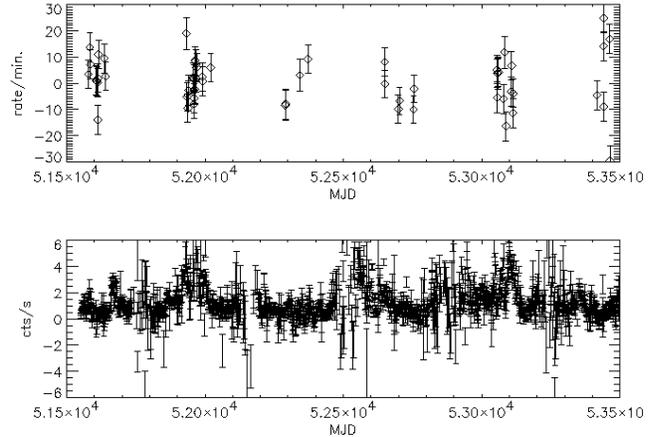}
\caption{Average $\gamma$-ray rate per night from 2000 to 2005 from PACT (upper panel) 
along with daily average X-ray photon rate obtained by ASM (All Sky Monitor) on board 
RXTE for Mkn 421. PACT data points include statistical as well as systematic errors.}
\label{fig:6}
\end{figure}

\begin{figure}
\centering
  \includegraphics[width=0.4\textwidth]{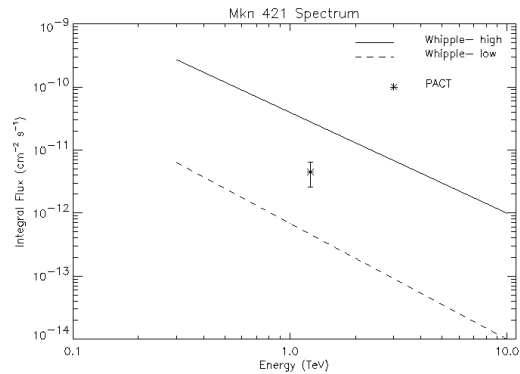}
\caption{Integral energy spectrum of Mkn421. The data point(asterisk) with 
 error bar represents the time averaged integral flux of Mkn 421 obtained 
 from PACT observations during 2000 to 2005. Dashed line represents the 
 flux measured by Whipple during a low activity state and solid line 
 represents their flux measurement during 2001 flaring state.}
\label{fig:7}
\end{figure}

   Most of our observations pertain to Mkn 421.  
  The upper panel of figure \ref{fig:6} shows the daily average of TeV 
 $\gamma$-ray rate as obtained from PACT for Mkn 421. In the lower panel of
 this figure  the daily average of X-ray photon rate as obtained by ASM 
 (All Sky Monitor) on board RXTE is shown for comparison. 
 PACT sensitivity is such that to have resonable
 signal to noise ratio we need to add data from several nights. Therefore we
 have obtained the time averaged flux of TeV $\gamma$-rays
 from Mkn421 by combining all data obtained using PACT during 2000-2005. This is
 found to be 4.5$\pm$1.9 $\times$10$^{-12}$ photons cm$^{-2}$ s$^{-1}$ above
 1.2 {\it TeV}. 
 Since Mkn 421 is observed at an average angle of about 30$^\circ$ with 
 respect to zenith, energy threshold of PACT for these observations
 corresponds to 1.2 {\it TeV}. In terms of Crab units this flux is
 0.3 units. The overall significance is low due to systematic errors
 and is 2.3 $\sigma$.  
  This integral flux from Mkn 421  as obtained by PACT
 is shown in figure \ref{fig:7}.  The flux of $\gamma$-rays
 during the most intense flare and lowest activity state as obtained by 
 Whipple group is also shown in this figure. The dotted line 
 represents the quiescent flux measured in 1995 \cite{Ref10} and solid line represents 
 the flux measured during a flaring state in 2001\cite{Ref11}. During the period of our
 observations, there were a few flares in 2000, 2001 and 2004. So our time 
 averaged flux is expected to be between the two extreme limits given by 
 Whipple and other Groups.
 
\begin{figure}
\centering
  \includegraphics[width=0.4\textwidth]{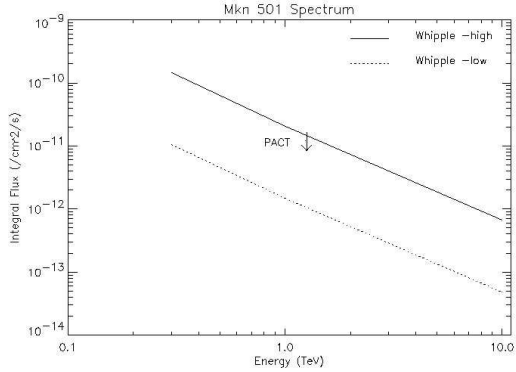}
\caption{ Integral energy spectrum of Mkn 501. The down arrow represents the 3$\sigma$
upper limit on the flux of TeV $\gamma$-rays from this source obtained by PACT. Dotted
line represents the flux measured by Whipple during a low activity state and solid line
 represents their flux measurement during 1997 flaring state.}
\label{fig:8}
\end{figure}

\begin{figure}
\centering
  \includegraphics[width=0.4\textwidth]{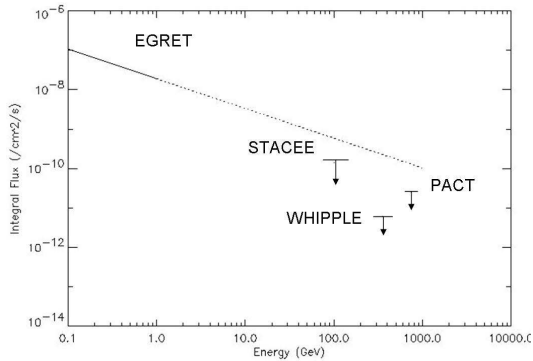}
\caption{ Upper limits on the flux of GeV/TeV $\gamma$-rays from ON231 as
obtained by Whipple, STACEE and PACT groups. EGRET spectrum is shown by solid line,
which is extended upto TeV energies using dotted line.}
\label{fig:9}
\end{figure}

 We have not seen any significant $\gamma$-ray flux from other blazars
 we have observed (Mkn501, 1ES1426+428 and ON231). Earlier, in 1997, a huge 
 flare was detected from Mkn501 \cite{Ref12,Ref13,Ref14}. But during our observation 
 period this source was in low state. It was much weaker than 
 Crab Nebula. We have estimated 3$\sigma$ upper limit on $\gamma$-ray flux 
 from Mkn 501, as 1.22$\times$10$^{-11}$ photons cm$^{-2}$ s$^{-1}$
 ($\geq$ 1.2 TeV). Average zenith angle for these observations was about
 30$^\circ$. This upper limit corresponds to 0.75 Crabs. This upper limit  
 is shown in figure \ref{fig:8}. The dotted line represents the
 quiescent flux measured in 1995 \cite{Ref10} and solid line represents the
 flux measured during the flaring state in 1997 by the Whipple group \cite{Ref13}.
 Blazar 1ES1426+428 is a distant object at z of 0.129. 
 Whipple and HEGRA groups have detected TeV $\gamma$-rays from this source with 
 long duration observations \cite{Ref15,Ref16}. 
 PACT is less sensitive than Whipple telescope. Also, we do not have very 
 long coverage for this source. We estimate  3$\sigma$ upper limit on 
 $\gamma$-ray flux for 1ES1426+428 to be 
 1.34$\times$10$^{-11}$ photons cm$^{-2}$ s$^{-1}$ ($\geq$ 1.2 TeV). Here
 energy threshold is in accordance with average zenith angle of 
 30$^\circ$ during these observations and upper limit corresponds to about 
 0.8 Crab flux.
 ON231 is a LBL type blazar and till date TeV $\gamma$-rays are detected
 only from HBL blazars. But, observations by EGRET on board 
 Compton Gamma Ray Observatory (CGRO) have shown a hard power law energy 
 spectrum  (photon spectral index $\alpha$ = 1.73 $\pm$ 0.18) extending upto 
 about 10 GeV with no sign of any cutoff. Because of its hard spectrum 
 it was thought that ON231 may be detected at higher energies and 
 hence is a potential TeV $\gamma$-ray source. GeV/TeV $\gamma$-rays 
 have not yet been detected from ON231 so far \cite{Ref17,Ref18}.
 ON231 was observed by PACT at an average zenith angle of about 10$^\circ$ 
 with energy threshold about 800 GeV.
 We have estimated 3$\sigma$ upper limit on $\gamma$-ray flux from this 
 source as 2.50$\times$10$^{-11}$ photons cm$^{-2}$ s$^{-1}$ ($\geq$ 800 GeV)
 which corresponds to 0.83 Crab units
 and is shown in figure \ref{fig:9} along with other results.

\subsection{Spectral Energy Distribution of Mkn 421}
\label{sec:5}
 
   We have derived the spectral energy distribution of Mkn 421 during high 
 and low states. The former is based on contemporaneous radio, optical, X-ray 
 and $\gamma$-ray observations while the later is based on 
 contemporaneous radio, NIR, X-ray and $\gamma$-ray data. 

  Spectral analysis of X-ray data was done using XSPEC. Spectral data from both
 the data sets were fitted by cutoff power law with line of sight absorption. 
 Line of sight absorption was fixed to neutral hydrogen column density at
 1.38 $\times$ 10$^{20}$ {\it cm$^2$} \cite{Ref19}. The best fit photon 
 indices for 2001 and 2003 data are found to be 2.05$\pm$0.03 and 
 2.40$\pm$0.03 and cutoff energies are about 24.9$\pm$0.26 {\it keV} and 
 23.9$\pm$2.4 {\it keV} respectively. 

\begin{figure}
\centering
  \includegraphics[width=0.5\textwidth]{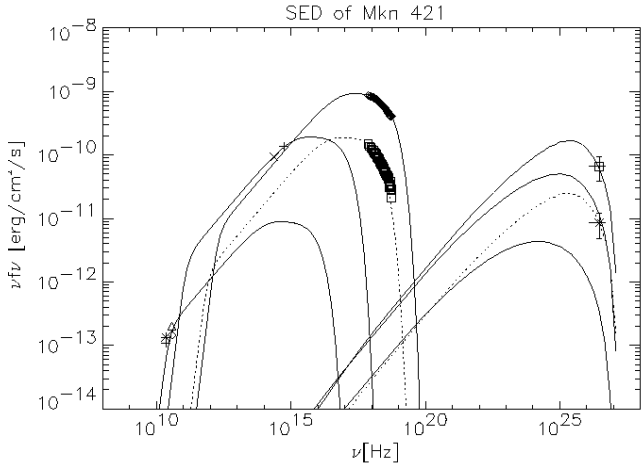}
\caption{SED of Mkn 421 for 2001 (flare state) and 2003 (quiescent state).
SSC fits for both these states are shown by solid and dotted lines respectively.
Three one-zone SSC models are used to fit X-ray, optical and radio data in each case.}
\label{fig:10}
\end{figure}

 Figure \ref{fig:10} shows spectral
 energy distributions (SEDs) obtained for Mkn 421 involving multiwaveband 
 data for 2001 and 2003.  We have used the time averaged
 flux of $\gamma$-rays for Mkn 421 in the SED plot of 2003. During the
 multiwavelength campaign in 2003  this source was at quiescent state as
 mentioned earlier. Mkn421 was at low state during most of the PACT observations
 from 2000 to 2005 and hence the time averaged flux is expected to be
 closer to the quiescent state flux of Mkn421. For 2001 flare we have selected
 PACT data overlapping with X-rays. After applying selection cuts we are
 left with 4.8 hours of data which is simultaneous with X-ray data and
 the corresponding $\gamma-$ray flux is 3.4$\pm$1.4 $\times$ 10$^{-11}$
 photons cm$^{-2}$ s$^{-1}$. 
 We have fitted the
 SEDs of both these states with a simple one-zone SSC model (for detailed 
 description of the code see Krawzynski et al. \cite{Ref20}) as shown in
 figure \ref{fig:10}: solid line for 2001 and dotted line for 2003.
 This model assumes 
 spherical blob of radius R and uniform magnetic field B, moving with respect 
 to the observer with the Doppler Factor $\delta$, which is filled with a 
 homogeneous non-thermal electron population. 
 Fit to SED of Mkn 421 spanning X-ray and $\gamma$-ray energies using SSC
 model has been attempted by number of authors for low as well as high state of
 the source. The fitted parameters broadly fall under two categories : some 
 prefered larger Doppler factor $\delta$ of about 50 (\cite{Ref(Konopelko)},
 \cite{Ref(Rebillot)} and \cite{Ref(Albert)}) while some others used lower 
 value of $\delta$ in the range
 10-20 \cite{Ref21}. We have tried both these cases ($\delta$=50 and 14) 
 and the
 fits for $\delta$=50 are shown in figure \ref{fig:10}. There is substantial change 
 in magnetic field (B) and radius of blob (R) in the two states.
 We obtained B(magnetic field)=0.20 G, R(radius)=2.3$\times$10$^{13}$ m and 
 w(electron density)=0.03 erg/cm$^3$ for the low state in 2003 with $\delta$=50.
 For the high state in 2001 the respective values are 0.57 G, 1.3$\times$10$^{13}$ m
 and 0.1 erg/cm$^3$. For $\delta$=14 the values for B, R and w are 0.28 G, 
 1$\times$10$^{14}$
 and 0.01 erg/cm$^3$ for the low state and 0.4 G, 9$\times$10$^{13}$ m and 
 0.04 erg/cm$^3$ for the high state. However it is also possible to fit the 
 SEDs by varying R and B but keeping $\delta$ same.     

   Synchroton peak in the SED for 2001 is
 located at higher energies compared to that of 2003 as shown in figure
 \ref{fig:10}, implying spectrum hardens as flux increases. Even though one 
 zone SSC model fits high energy emission nicely (for 2001 and 2003) it fails 
 to take into account radio and optical fluxes. If we assume additional 
 electron populations, as suggested by Krawzynski et al. for 
 1ES 1959+650, responsible for the low energy emission, then the SED's of 
 these components could also be fitted. An extensive study of Mkn421 was 
 carried out by Blazejowski et al. \cite{Ref21} for the period 2003-2004 
 involving radio, optical, X-ray and $\gamma$-ray data. There is good 
 agreement between SSC parameters obtained by them and here using $\delta$ $\sim$ 14.
 
\section{High Altitude GAmma Ray observatory}
\label{sec:6}

\begin{figure}
\centering
  \includegraphics[width=0.5\textwidth]{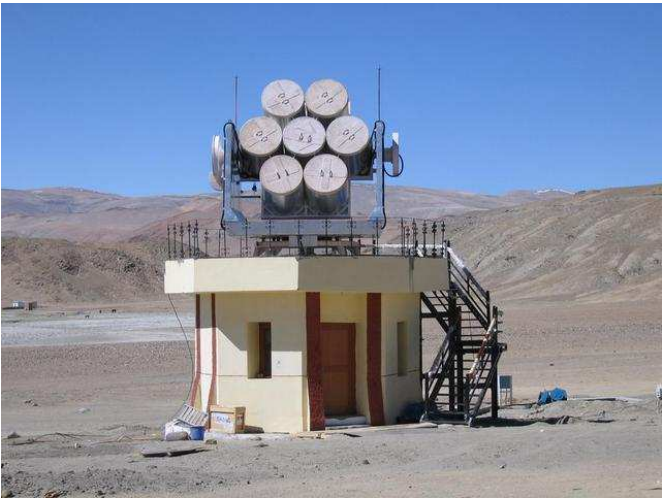}
\caption{Photograph of one of the telescope installed at Hanle}
\label{fig:12}
\end{figure}

   A 7 telescope array, called High Altitude GAmma Ray observatory (HAGAR),
 is being built at Hanle in the Himalayas, at an altitude of about 4.3 km, 
 above mean sea level~\cite{Ref22}. HAGAR is based on {\it wavefront 
 sampling technique} 
 like PACT. These 7 telescopes will be in the form of a hexagon with an
 intertelescope spacing of 50 m.  
 Each telescope has 7 para-axially mounted parabolic 
 mirrors of diameter $\sim$0.9 m with a photomultiplier tube at the 
 focus of each mirror. The atmospheric attenuation of Cherenkov photons at 
 Hanle altitude is $\sim$ 14\% as compared to $\sim$ 50\% at sea level. The 
 Cherenkov photon density near the shower core at Hanle is higher by a factor 
 4-5 compared to that at the sea level for showers of same energy. 
 These features effectively reduce the energy threshold of HAGAR which 
 is estimated to be $\sim$ 60 GeV for vertically incident $\gamma$-ray showers. 
 Sensitivity of HAGAR would be such that it will detect the Crab at 
 5$\sigma$ level without any hadron rejection in $\sim$ 2 hours. Figure \ref{fig:12}
 shows the photograph of the first telescope commissioned in June, 2005. 
 Comissioning of remaining
 telescopes is underway. All  7 telescopes are expected to be operational 
 by middle of 2007.

\section{Conclusions}
\label{sec:7}

Out of 4 blazars which we have observed using PACT only Mkn 421 was reported
(by other experiments) to be in flaring state on few occassions between 2000-2005. 
We have estimated average integral flux for this source by combining all the 
data from 2000 to 2005. For others, Mkn 501, 1ES1426+428 and ON231 we have 
given 3$\sigma$ upper limit on $\gamma$-ray flux.  

 We have found from the study of SEDs of Mkn421 that X-ray and $\gamma$-ray emissions
during high and low states are correlated. Synchrotron peak of SED of 2001 is at 
higher energy 
compared to that of 2003, suggesting that peak shifts towards higher energy as flux 
increases. There are significant changes in SSC parameters for these two 
data sets at X-ray and $\gamma$-ray energies. One zone SSC model can not 
fit all data, introduction of additional zone improves the fit at lower
energies. But there is almost no change for SSC parameters at lower energies
during quiescent and flaring state of Mkn421.

\begin{acknowledgements}
 
 We are thankful to Dr. Talvikki Hovatta for providing us the published
radio data. We gratefully acknowledge the use of RXTE data from the public
archive of GSFC/NASA. 
We thank Prof. P. N. Bhat, S.S. Upadhya, K.S. Gothe, B.K. Nagesh, 
S.K. Rao, M.S. Pose, P.V. Sudershanan, S. Sharma, K.K. Rao, 
A.J. Stanislaus, P.N. Purohit, A.I. D'Souza, J. Francis, and 
B.L.V. Murthy for their support during construction, maintenance of 
PACT and observations.  

\end{acknowledgements}

\end{document}